\documentclass[a4paper,11pt]{article}
\pdfoutput=1
\usepackage{jheppub}

\usepackage{graphicx}
\usepackage{placeins}
\usepackage{url}
\usepackage{multirow}
\usepackage{amssymb}
\usepackage{amsmath}
\usepackage{xspace}
\usepackage{booktabs}
\usepackage{xcolor}
\usepackage{cleveref}
\usepackage[symbol]{footmisc}

\DeclareMathOperator*{\argmax}{arg\,max}

\usepackage{algorithm}
\usepackage[noend]{algpseudocode}

\newcommand{\bn}[1]{{\color{red}[BN: #1]}}

\begin{document}

\title{Stabilizing Neural Likelihood Ratio Estimation}

\affiliation[a]{Physics Division, Lawrence Berkeley National Laboratory, Berkeley, CA 94720, USA}
\affiliation[b]{Department of Physics, University of California, Berkeley, CA 94720, USA}
\affiliation[c]{NERSC, Lawrence Berkeley National Laboratory, Berkeley, CA 94720, USA}
\affiliation[d]{Berkeley Institute for Data Science, University of California, Berkeley, CA 94720, USA}

\author[a]{Fernando Torales Acosta,}
\author[a,b]{Tanvi Wamorkar,}
\author[c]{Vinicius Mikuni,}
\author[a,d]{and Benjamin Nachman}

\abstract{
Likelihood ratios are used for a variety of applications in particle physics data analysis, including parameter estimation, unfolding, and anomaly detection.  When the data are high-dimensional, neural networks provide an effective tools for approximating these ratios.  However, neural network training has an inherent stochasticity that limits their precision.  A widely-used approach to reduce these fluctuations is to train many times and average the output (ensembling).  We explore different approaches to ensembling and pretraining neural networks for stabilizing likelihood ratio estimation.  For numerical studies focus on unbinned unfolding with OmniFold, as it requires many likelihood ratio estimations.  We find that ensembling approaches that aggregate the models at step 1, before pushing the weights to step 2 improve both bias and variance of final results. Variance can be further improved by pre-training, however at the cost increasing bias.

}

\maketitle

\section{Introduction}

Simulations are critical tools for inference in particle physics.  These synthetic datasets connect the fundamental theory with observables to enable a wide range of statistical analyses.  Many of these analyses require the simulation to match the data in certain regions of phase space or to be smoothly dependent on model parameters.  These tasks can be accomplished by reweighting a simulated dataset using likelihood ratios between the target and initial datasets.  Traditionally, likelihood ratios have been estimated by taking ratios of histograms.  Modern machine learning tools have enabled unbinned, high-dimensional inference.  By avoiding dimensionality reduction and discretization, these new tools can improve precision and accuracy by using (much) more information from the data.  In particular, neural simulation-based inference (nSBI)~\cite{Cranmer_2020} scaffolds simulations with deep learning for optimal data analysis.

Training deep neural networks has a random element that poses a significant challenge for nSBI.  For most applications involving neural networks, (sub) percent-level variations in the outputs have little impact on the physics sensitivity.  For nSBI, these small fluctuations add up and can reduce the precision of a measurement.  For example, if there is a 1\% variation in the nominal result and the measurement process is repeated $N$ times with variations due to systematic uncertainties, then there is an apparent $\sqrt{N}\%$ uncertainty even if the systematic shifts do nothing.  This artificial uncertainty can quickly become comparable or larger then the leading real uncertainty, limiting the utility of the result.

There are three approaches to mitigating the fluctuations from neural network training: hyperparameter optimization (HPO), optimizer selection, and ensembling.  Activation and loss functions have been well-studied~\cite{Cranmer:2015bka, Kong:2022rnd, Brehmer_2018, Brehmer_2018_long, D_Agnolo_2019, Nachman_2021, Stoye:2018ovl, 2019arXiv191100405M, Jeffrey:2023stk,Rizvi:2023mws} and recent proposals for new optimizers have explored likelihood ratio stability~\cite{DeLuca:2025ruv}.  The most widely-used tool in all experimental measurements using nSBI is ensembling: running the training many times and averaging the result per event~\cite{H1:2021wkz,H1prelim-22-031,H1:2023fzk,H1:2024mox,LHCb:2022rky,Song:2023sxb,Pani:2024mgy,ATLAS:2024xxl,ATLAS:2025qtv,CMS-PAS-SMP-23-008,ATLAS:2024jry}.  Despite its widespread use, we are not aware of a dedicated study of ensembling methods.

In this paper, we systematically study different ensembling methods.  The main questions are how to control the randomness across trainings and then when and how to combine the results.  References~\cite{Mikuni:2024qsr,Mikuni:2025tar} showed how pre-training can accelerate likelihood ratio estimation and improve the accuracy, but we are mostly interested in the complementary question of precision.  For numerical results, we focus on the nSBI task of unfolding, which is the statistical removal of detector effects to perform differential cross section measurements.  There are many proposals for unbinned unfolding methods~\cite{Arratia:2021otl}; we focus on the OmniFold approach~\cite{Andreassen:2019cjw,Andreassen:2021zzk} as it is the one that has been used for all experimental results so far~\cite{H1:2021wkz,H1prelim-22-031,H1:2023fzk,H1:2024mox,LHCb:2022rky,Song:2023sxb,Pani:2024mgy,ATLAS:2024xxl,ATLAS:2025qtv,CMS-PAS-SMP-23-008} and likelihood-ratio estimation is a core task of the method.  In particular, OmniFold is an iterative two-step procedure where a likelihood ratio is estimated at each step of every iteration.  The second step likelihood ratio mitigates prior dependence and is also used by other algorithms~\cite{Backes:2022sph}.  We explore ensembling at every step of every iteration (`step') or ensembling at the level of the full algorithm (`parallel').  Additionally, we study the interplay between ensembling and pre-training, which also has the potential to reduce the variance.  While the former applies to iterative algorithms, the latter is relevant for all likelihood-ratio estimation tasks.

This paper is organized as follows. Section~\ref{sec:unfolding} briefly describes OmniFold and the baseline approach to neural likelihood ratio estimation.  The dataset used for numerical results -- an all-particles differential cross section measurement in high-$Q^2$ deep inelastic scattering with a simulated H1 detector at the HERA collider -- is introduced in Sec.~\ref{MC}.  Section~\ref{sec:methods} provides details of the ensembling and pre-training strategies.  Numerical results are presented in Sec.~\ref{results} and the paper ends with conclusions and outlook in Sec.~\ref{sec:conc}.

\section{Neural Likelihood Ratio-based Unfolding}
\label{sec:unfolding}

One widely-used method for estimating likelihood ratios with neural networks is repurposing classifiers.  The optimal binary classification score is any monotonic function of the likelihood ratio~\cite{neyman1933problem}.  Thus, when the monotonic function is known, a binary classifier can be converted into a likelihood-ratio estimator.  In the continuum limit, functional optimization can connect the learned statistic with a given loss function~\cite{Nachman:2021yvi}.  It is well-known in particle physics that for the mean-squared error or binary cross entropy loss~\cite{Cranmer:2015bka}, the classifier output $f(x)$ approximates $p(A|x)$ where $x\in\mathbb{R}^N$ are the features and $A$ is the class assigned a numeric label of 1 (while class $B$ is labeled $0$).  For equal numbers of examples from both classes used in the training,

\begin{align}
\label{eq:likelihoodratiotrick}
    \frac{f(x)}{1-f(x)}\approx\frac{p(x|A)}{p(x|B)}\,.
\end{align}
Due to the ubiquity of this approximation, we will use the binary cross-entropy loss and Eq.~\ref{eq:likelihoodratiotrick} when estimating likelihood ratios.

Unfolding is a two-level estimation problem with particle-level $z\in\mathbb{R}^N$ and detector-level $x\in\mathbb{R}^M$.  Data events are observed at detector-level and synthetic events are generated in $(z,x)$ pairs\footnote{For acceptance/efficiency effects, $x$ or $z$ is set to a default value if the event at detector- or particle-level fails the event selection.}.  OmniFold is an unfolding algorithm based on likelihood ratios.  First, $\omega(x)\equiv p_\text{data}(x)/p_\text{sim.}(x)$ is estimated using Eq.~\ref{eq:likelihoodratiotrick}.  This induces a new simulated distribution where each event $i$ is weighted by $\omega(x_i)$.  The second step of OmniFold estimates a likelihood ratio between the induced particle-level spectrum and the initial particle-level spectrum.  Symbolically, the new weight $\nu(z)=f(z)/(1-f(z))$, where

\begin{align}
\label{eq:step2}
    f(z)=\argmax_g\sum_\text{sim.}\omega(x_i)\log(g(z_i))+ \sum_\text{sim.}\log(1-g(z_i))\,.
\end{align}
This second step is necessary because $\omega$ is a function of the detector-level phase space while the final result should be a function of the particle-level phase space only.  The output of OmniFold is the events $z_i$ with weights $\nu(z_i)$.  Any statistic, histogram, etc. can then be computed from these data.  Typically, the entire process is repeated by pushing the weights $\nu(z_i)$ back to the detector level: $\omega_{n+1}(x)=h_{n+1}(x)/(1-h_{n+1}(x))$, where

\begin{align}
    h_{n+1}(x)=\argmax_g\sum_\text{data.}\log(g(x_i))+ \sum_\text{sim.}\nu_n(z_i)\log(1-g(x_i))\,.
\end{align}
with $\nu_0(z)=1$ and $\omega_1(x)=\omega(x)$.  Subsequently, the second step can be repeated by replacing $\omega$ in Eq.~\ref{eq:step2} with $\omega_{n+1}$ to learn $\nu_{n+1}$. This iterating removes the prior dependence and moves the solution closer to the maximum likelihood estimator.  Truncating early is a form of regularization, typically with 3-5 iterations.

For our applications, $z$ is the full set of final state particles from a collision and $x$ is the complete set of reconstructed particles.  These are both variable-length point clouds where there is no natural notion of order (permutation invariance).  In order to accommodate these properties of the inputs, all of the neural networks are built on the transformer architecture~\cite{NIPS2017_3f5ee243}.  The particular architecture is the point-edge transformer (PET)~\cite{Mikuni:2025tar,Mikuni:2024qsr}, designed for sets of particles, where each particle has a four-vector and additional properties like electric charge.  The PET model combines the scalability of transformer models and encodes local information using graph neural networks. In particular, a local embedding is created for each (reconstructed) particle using the $k$-nearest neighbors of each particle in $\eta-\phi$ space, where $k$ is set to 5. The representation of each (reconstructed) particle is then updated based on the local embedding information and used as an input to multiple transformer layers. The classification output is encoded using a trainable classification token~\cite{dosovitskiy2020image}, which summarizes the information from the entire point cloud and returns a binary output used for the classification task.  The stochasticity from retraining the network is mostly due to the randomness in the initial values of the weights and biases of the model.

All hyperparameters of the model, including number of layers, number of neighbors, and layer sizes were optimized by running multiple trainings and tracking the value of the validation loss of the classification using the datasets described in the next section. Since the entire procedure is computationally expensive, we only considered options that lead to acceptable running times, but we did check that using more complex models did not lead to noticeable improvements in the validation loss. All networks are implemented in \textsc{TensorFlow}~\cite{tensorflow} and \textsc{Keras}~\cite{keras} and optimized using \textsc{Lion}~\cite{chen2024symbolic} with a learning rate of 2$e^{-5}$ and early stopping with a patience of 10.  All inputs are scaled so that each input feature has mean zero and unit standard deviation at reconstruction level.

\section{Datasets and Metrics}
\label{MC}

High $Q^2$ deep-inelastic scattering events from the H1 detector at HERA are used as representative collision events for testing different approaches.  The simulated samples we employ are the same as those used in Refs.~\cite{Arratia:2021tsq,Arratia:2022wny,Long:2023mrj} and whose properties are briefly summarized below.

Particle-level simulations are performed with Djangoh 1.4~\cite{Charchula:1994kf} and Rapgap 3.1~\cite{Jung:1993gf}.  In our experiments, Djangoh plays the role of `data' while Rapgap is used as the `simulation' to perform the unfolding.  Particle-level events are then passed through a detailed simulation of the H1 detector built on \textsc{Geant}3~\cite{Brun:1987ma} and the output is reconstructed using an energy flow algorithm~\cite{energyflowthesis,energyflowthesis2,energyflowthesis3}.  Particle-level events are required to have $Q^2 > 100$ GeV$^2$ and at least one particle in the event.  Detector-level events have a more stringent phase space selection, with $Q^{2}> 150$ GeV$^{2}$, inelasticity $0.08<y<0.7$, and particles with $p^\mathrm{part}_\mathrm{T} >$ 0.1 GeV and $-1.5 < \eta^\mathrm{part.} < 2.75$ are selected.  The more restrictive selection at reconstruction level is to ensure a high level of reconstructability and to minimize the extrapolation of acceptance effects outside the phase space of interest.

Each event is specified by a set of per-particle features as well as event-level features.  The per-particle features are transverse momentum $p_T^\text{part.}$, pseudorapidity $\eta^\text{part.}$, azimuthal angle $\phi$ and electric charge $C$.  The event-level quantities are the $Q^2$, $y$, and electron momentum $p_x^e, p_y^e$, and $p_z^e$.  To focus on regions of physical relevance, we present $\log_{10}(p_\mathrm{T}^\mathrm{part.})$, $\log_{10}p_\mathrm{T}^\mathrm{part.}/Q$, and $\eta^\mathrm{part.}-\eta^e$ as well as the electron pseudorapidity $\eta^e$. We also consider a number of derived observables, including those based on jets at particle level.   Jets are reconstructed using the $k_\mathrm{T}$ algorithm~\cite{Catani:1993hr,Ellis:1993tq} with a resolution parameter of $R=1.0$, run on unfolded-particles. Jets with transverse momentum $p_\mathrm{T}^\mathrm{jet}>$ 5 GeV are selected for analysis. $p_\mathrm{T}^\mathrm{jet}$ is the transverse-momentum of the reconstructed jet, $\eta^\mathrm{jet}$ is the pseudo-rapidity of the jet, and jet $\tau_{21}$ is the ratio of two $N$-subjettiness observables~\cite{Thaler:2011gf,Thaler:2010tr}: $\tau_1 / \tau_2$.  Another important observable is the difference in azimuthal angle between a given jet and the scattered lepton: $\Delta\phi^\text{jet}$ .


The goal is to apply the unfolding method from the previous section to estimate the particle-level Djangoh using pairs of particle-level/detector-level events from Rapgap.  We are interested in minimizing both the bias (often used as a closure test to set a bias uncertainty) and the variance from rerunning the entire unfolding procedure multiple times. We use two metrics for each observable, constructing bins after the unfolding to study the statistical properties of the results\footnote{The binning is not required and other metrics that are not based on histograms (such as distribution moments) could also be used.}. To investigate the bias, we use the relative mean-squared error (MSE):

\begin{equation}
    \label{eq:MSE}
    \mathrm{MSE_{rel.}} = \frac{\frac{1}{N_\mathrm{}} \sum_{i=1}^{N} (x_i - \hat{x}_i)^2}{\mu_x},
\end{equation}
where $N$ is the number of full passes for any given method, $x$ are the truth histogram values (Djangoh), $\mu_{x}$ is the mean of the truth histogram values, and $\hat{x}$ are the unfolded values. To investigate model variance, we use the relative standard deviation:

\begin{equation}
\label{eq:stdev}
\sigma_\mathrm{rel.}^2 = \frac{\sqrt{\frac{1}{N} \sum_{i=1}^{N} (\hat{x}_i - \mu_{\hat{x}})^2}}{\mu_{\hat{x}}},
\end{equation}
where $\mu_{\hat{x}}$ is the mean of the unfolded values.
The metrics are calculated from a set of $N=10$ full passes of each method.


\section{Methods}
\label{sec:methods}

\subsection{Ensembling}
\label{ensembling}
\label {ensembling}

We explore two ensembling strategies: repeat the entire procedure $N_\text{Ensembles}$ times (parallel ensembling) then average the ensembles, or repeat each component of neural network training $N_\text{Ensembles}$ times, average over the ensembles for that component -- or step, in the case of OmniFold -- then push those averaged weights to the next step (step ensembling).  For OmniFold, parallel ensembling leads to repeating a full pass through the method (step 1 and step 2 over a number of iterations) many times while step ensembling leads to repeating each step of each iteration many times, ensembling as the algorithm progresses.  For parallel ensembling, each pass of the full method is independent of every other pass and thus the entire process is easily parallelized.  A full pass of the parallel ensembling procedure is outlined in Algorithm 1.  Our default approach to combining weights is to take the average, but we also consider the median and a truncated mean (mean without the highest/lowest values).

    \begin{algorithm}[H]

\caption{Parallel-Ensembling (1 Full Pass)}
\begin{algorithmic}[1]
\For{$N_{\text{Ensembles}}$}
    \For{$N_{\text{Iterations}}$}
        \For{Step in [step 1, step 2]}
            \State Train Model
            \State Compute and apply likelihood ratio estimate as in Sec.~\ref{sec:unfolding}
        \EndFor
    \EndFor
\EndFor
\State Combine (e.g. by averaging) $N_\mathrm{Ensemble}$ Models
\end{algorithmic}
\end{algorithm}

On the other hand, step ensembling is not trivially parallizable because the networks trained at a given iteration depend on the networks trained at the previous iterations.  However, the hypothesis is that each step of every iteration is now stabilized, so the overall result should be more stable.  A full pass of the step ensembling procedure is outlined in Algorithm 2.
\begin{algorithm}[H]
\caption{Step-Ensembling (1 Full Pass)}
\begin{algorithmic}[1]

\For{$N_{\text{Iterations}}$}
    \For{Step in [step 1, step 2]}
        \For{$N_{\text{Ensembles}}$}
            \State Train Model
        \EndFor
        \State Combine (e.g. by averaging) $N_\mathrm{Ensemble}$ Models
        \State Compute and apply likelihood ratio estimate as in Sec.~\ref{sec:unfolding}
    \EndFor
\EndFor
\end{algorithmic}
\end{algorithm}

Algorithms 1 and 2 describe one full pass of the two ensembling procedures performed on OmniFold, each incorporating 5 iterations and 5 ensembles.  The entire pass is then repeated 10 times to compute the metrics introduced in the previous section.

\subsection{Pre-Training}
\label{sec:pretrain}

As the main source of stochasticity in the likelihood-ratio estimation is from neural network model initialization, strategies for setting the starting weights and biases may be particularly useful.     
%
%
One approach is to pre-train the model. Instead of training each model from scratch, each model starts from a targeted starting distribution of weights and biases. In this paper, we pre-train a classifier to discriminate between Rapgap and Djangoh\footnote{This is exactly step 1 of iteration 1 of OmniFold.  In practice, the pre-training would not use data directly, but since we only have two simulated samples, we use this setup here.}. This classifier has the same architecture as the step 1 and step 2 neural networks of OmniFold (see Sec.~\ref{sec:unfolding}) so that the network parameters can be loaded to train each part of the OmniFold procedure.

\begin{figure}[H]
    \centering
    \includegraphics[width=0.45\linewidth]{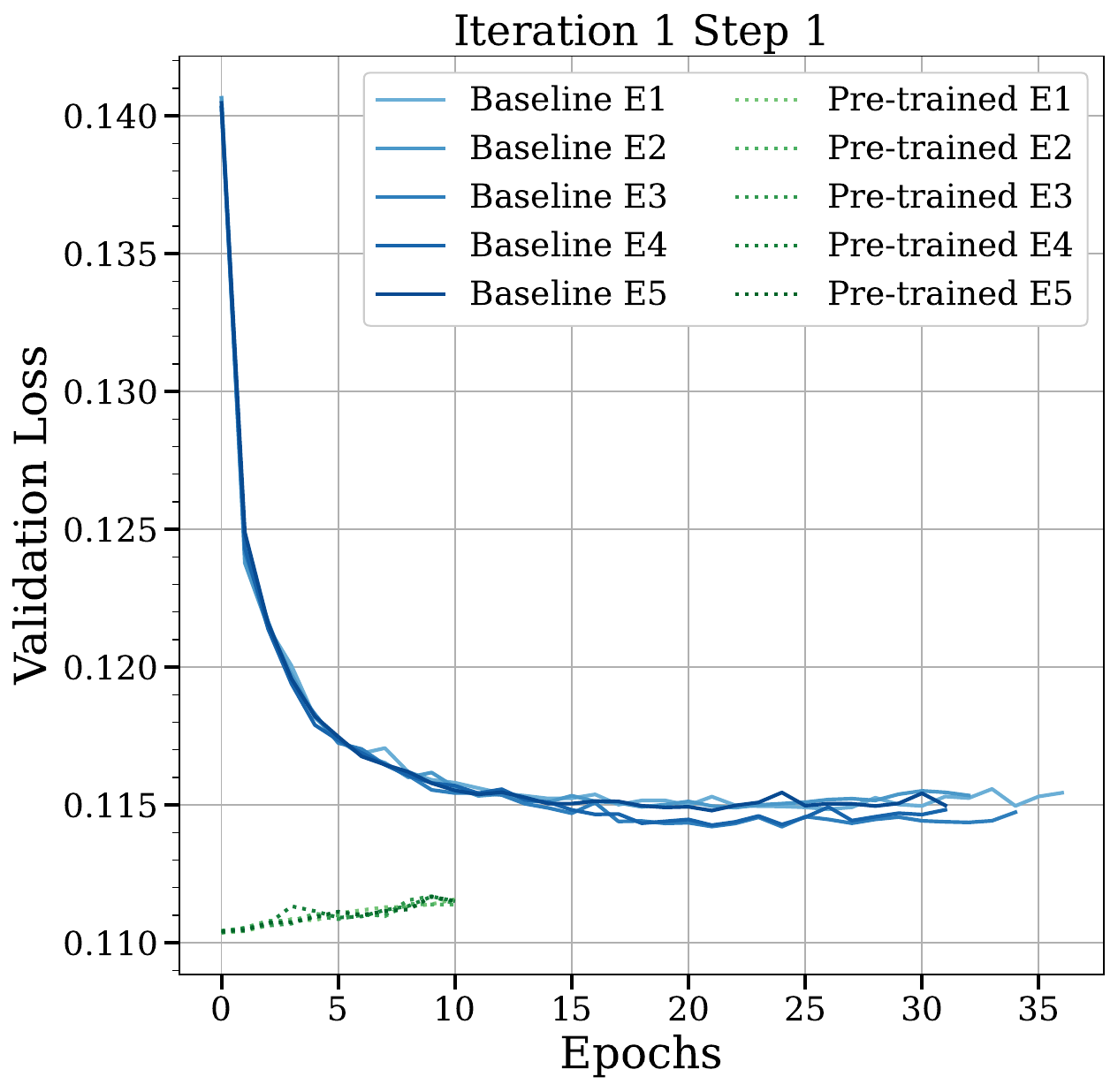}
    \includegraphics[width=0.45\linewidth]{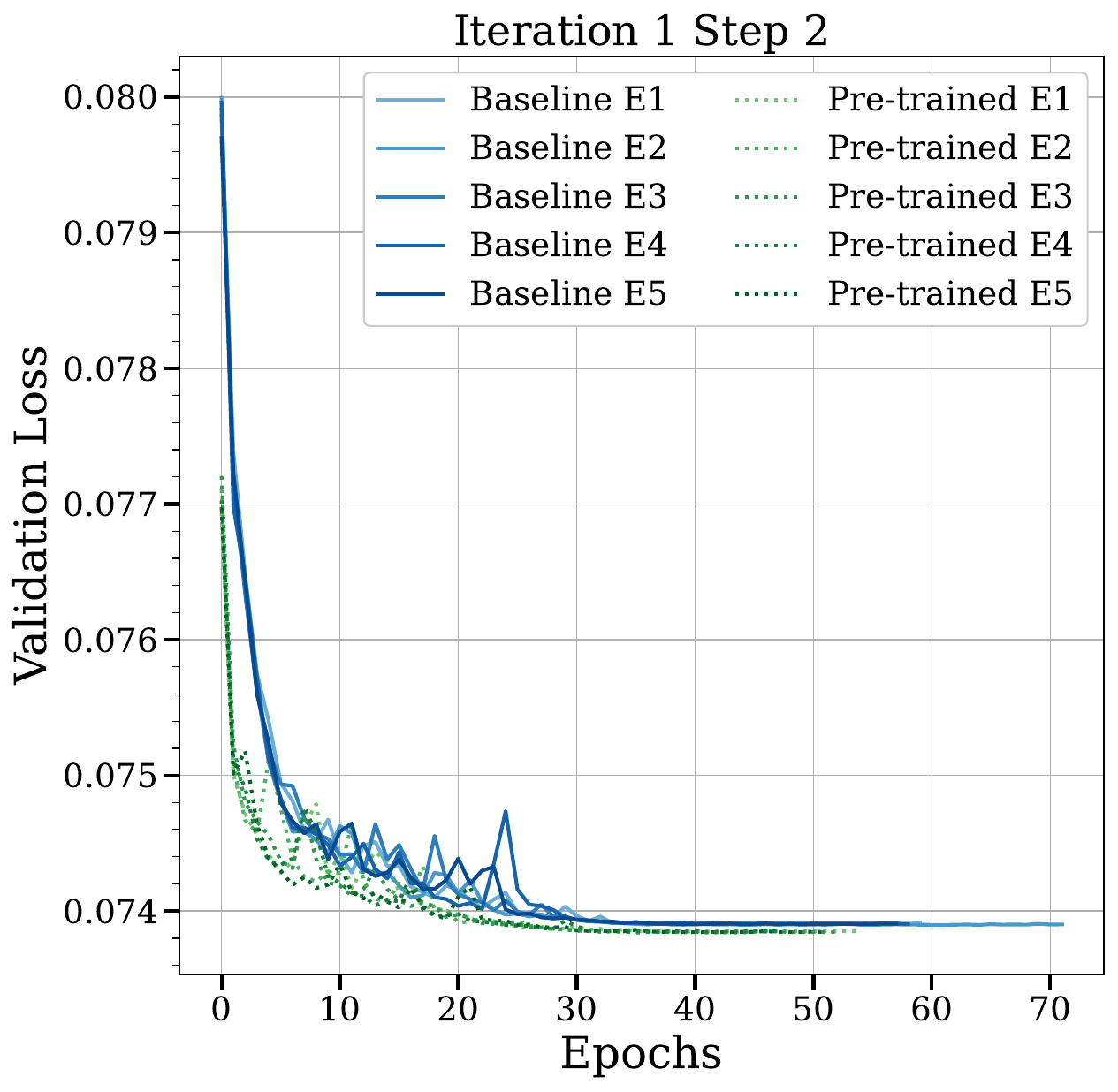}
    \caption{The validation loss for five passes of the first iteration of the OmniFold procedure using pre-trained classifiers (green) and baseline models, trained from scratch (green). All 5 ensembles (E1-E5) are shown for a full pass.
%
    %
    }
    \label{fig:val_loss}
\end{figure}

In earlier OmniFold measurements, each training began with a network initialized from random weights\footnote{Some studies have initialized the network at step $i+1$ with the one from step $i$.  Most of the training time and stochasticity is from the first iteration and so these differences do not have a big effect here.}. In the case of the H1 setting, the simulated samples introduced in Sec.~\ref{MC} are about two orders of magnitude larger than the corresponding data. This means that the unfolding precision was constrained by the data size and it is not necessarily to use all of the simulated data directly in the unfolding. For our pre-training option, we use 350k Rapgap and 350k Djangoh detector-level events. In contrast, the `data' contained 250,000 events, while the full Rapgapd and Djangoh samples available are 20 million events. A smaller number of events, as well as a higher learning rate of $1e^{-5}$ are used for pre-training (compared to the downstream training) to reduce the training time. This pre-trained model then serves as the foundation for both steps for all subsequent unfolding runs, improving both the quality and stability of the procedure. The model weights are updated after each iteration, and passed to the next iteration.
We demonstrate this by comparing the validation loss obtained from using the pre-trained model on the first OmniFold iteration with the loss observed when training a model from scratch, shown in Figure \ref{fig:val_loss}.  Since the pre-training task is the same as step 1 of iteration 1, the pre-training model stops after the patience of 10 epochs and the overall loss is lower and varies less.  The gains for step 2 are not as dramatic, but still visible by eye.

\section{Results}
\label{results}
\label{results}

First, we study the effect of different aggregation functions for combining the ensembles.  For illustration, we focus on the parallel ensembling approach, using the ensemble mean, truncated mean, and median. For the truncated mean, the lowest and highest model weights are pruned before taking the mean.  

\begin{figure}[H]
    \centering
    \includegraphics[width=0.99
    \linewidth]{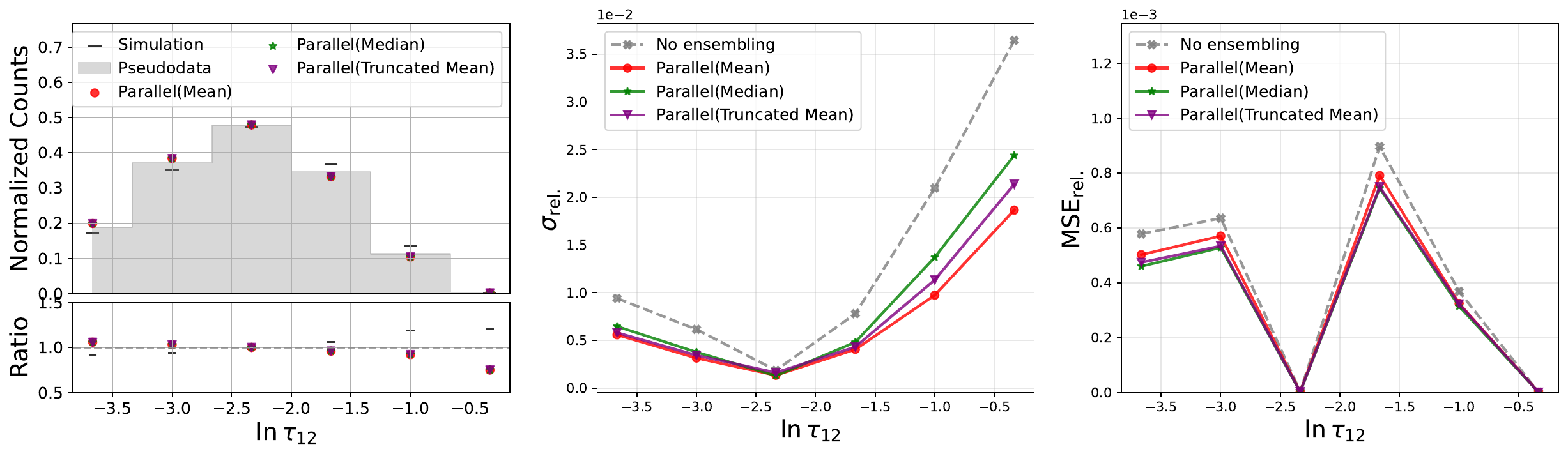}
    \caption{Comparison of jet $\tau_{12}$ unfolded with OmniFold using parallel ensembling method with mean (red), median (green) and truncated mean (purple) as a metric for aggregating the weights of $N_\text{Ensembles}$. The center panel shows the relative standard deviation of the 10 passes using the different metrics, and the right panel shows the relative MSE.
    }
    \label{fig:parallel_metrics}
\end{figure}

Figure \ref{fig:parallel_metrics} compares the three approaches for the $\tau_{21}$ observable.  This is one of many possible observables, but we found that the trends are similar: all three methods result in similar performance, with slightly better overall precision from the simple average.  This is therefore our baseline for all subsequent results.  We note that an advantage of parallel ensembling is that the aggregation function can be changed after the result, while it must be pre-specified for step ensembling.



\begin{figure}[H]
    \centering
    \includegraphics[width=0.95\linewidth]{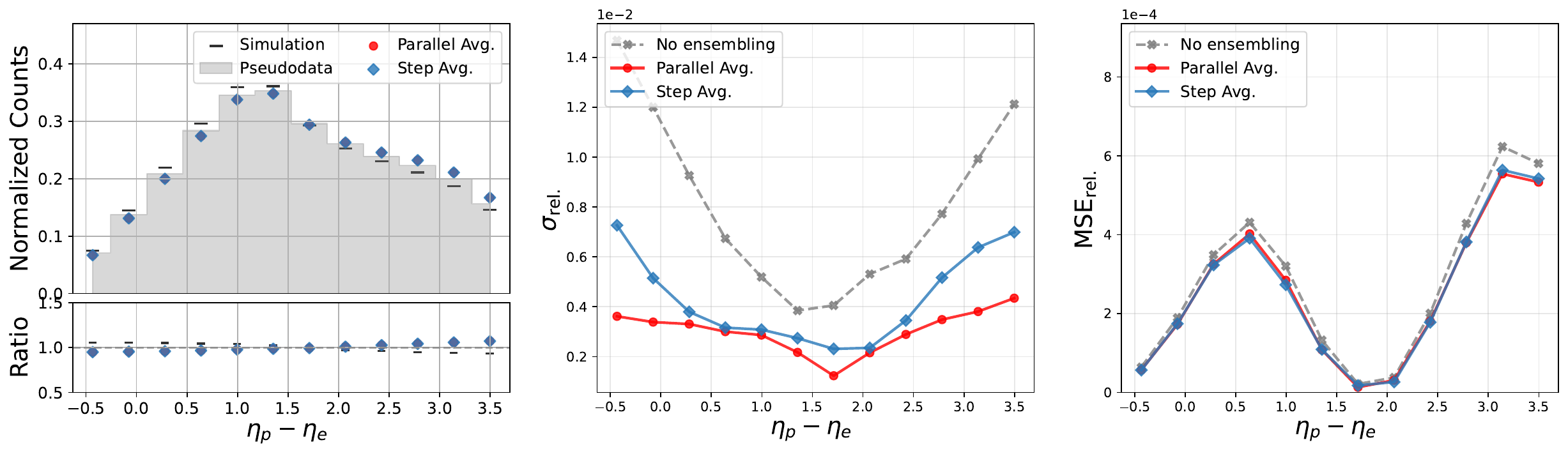}
    \includegraphics[width=0.95\linewidth]{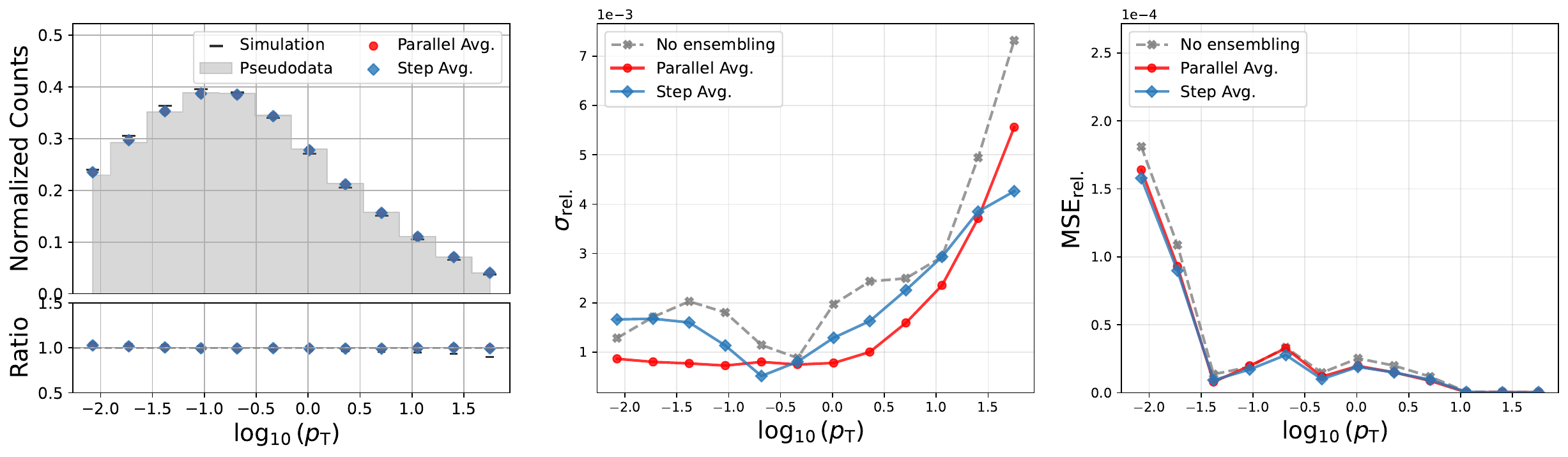}
    \includegraphics[width=0.95\linewidth]{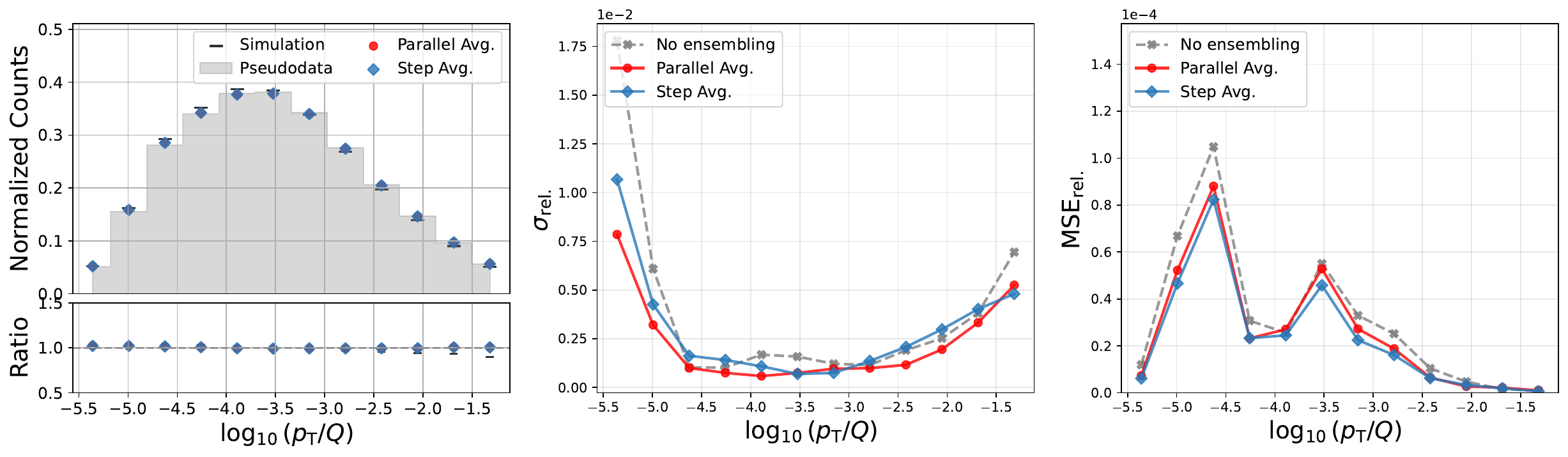}
    \caption{Comparison of relative particle pseudorapidity, $\eta_p-\eta_e$ (top), $\log_{10}(p_\mathrm{T}^\mathrm{part.})$ (middle), and $\log_{10}p_\mathrm{T}^\mathrm{part.}$ (bottom) for parallel and step ensembling methods with models trained from scratch. The middle column shows the relative standard deviation of the 10 passes, and the right column show the relative MSE.
    }
    \label{fig:part_obs}
\end{figure}

\begin{figure}[H]
    \centering
    \includegraphics[width=0.95\linewidth]{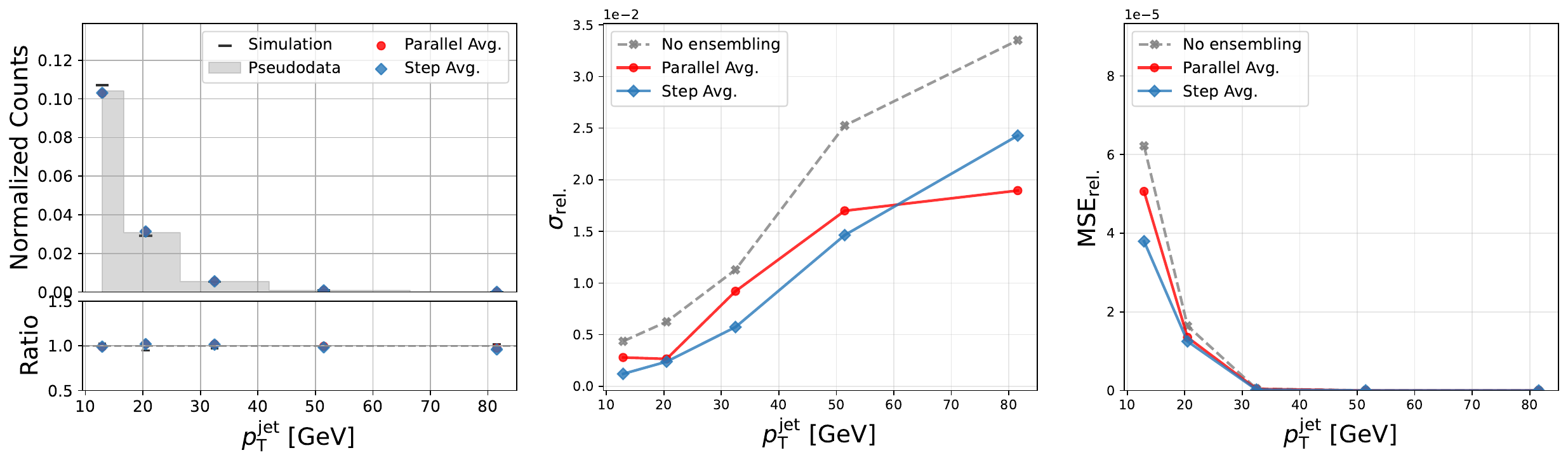}
    \includegraphics[width=0.95\linewidth]{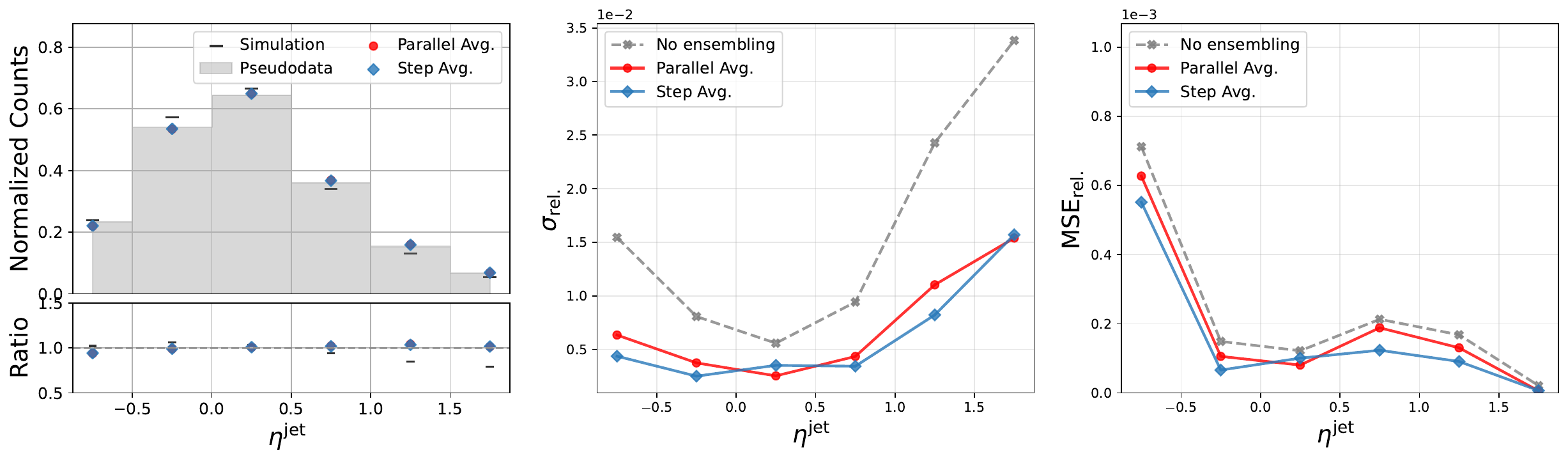}
    \includegraphics[width=0.95\linewidth]{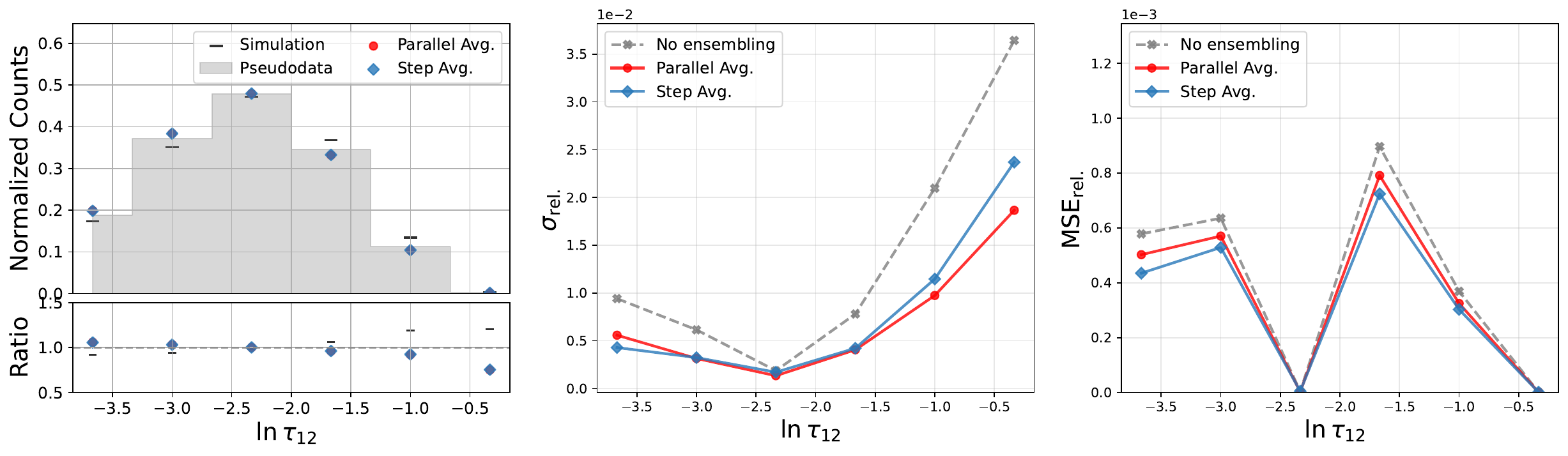}

    \caption{Comparison of $p_\mathrm{T}^\mathrm{jet}$ (top) $\eta^\mathrm{jet}$ (center) and jet $\tau_{21}$ (bottom) for parallel and step ensembling methods with models trained from scratch.  The middle column shows the relative standard deviation of the 10 passes, and the right column show the relative MSE.}
    \label{fig:jet_obs}
\end{figure}

\begin{figure}[H]
    \centering
    \includegraphics[width=0.95\linewidth]{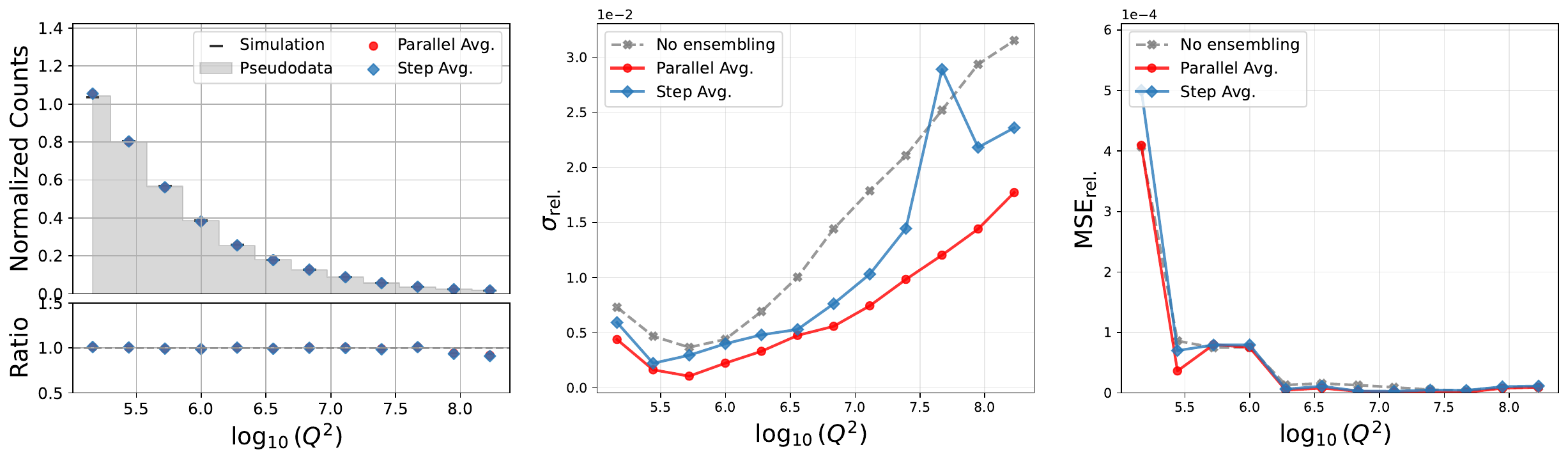}
    \includegraphics[width=0.95\linewidth]{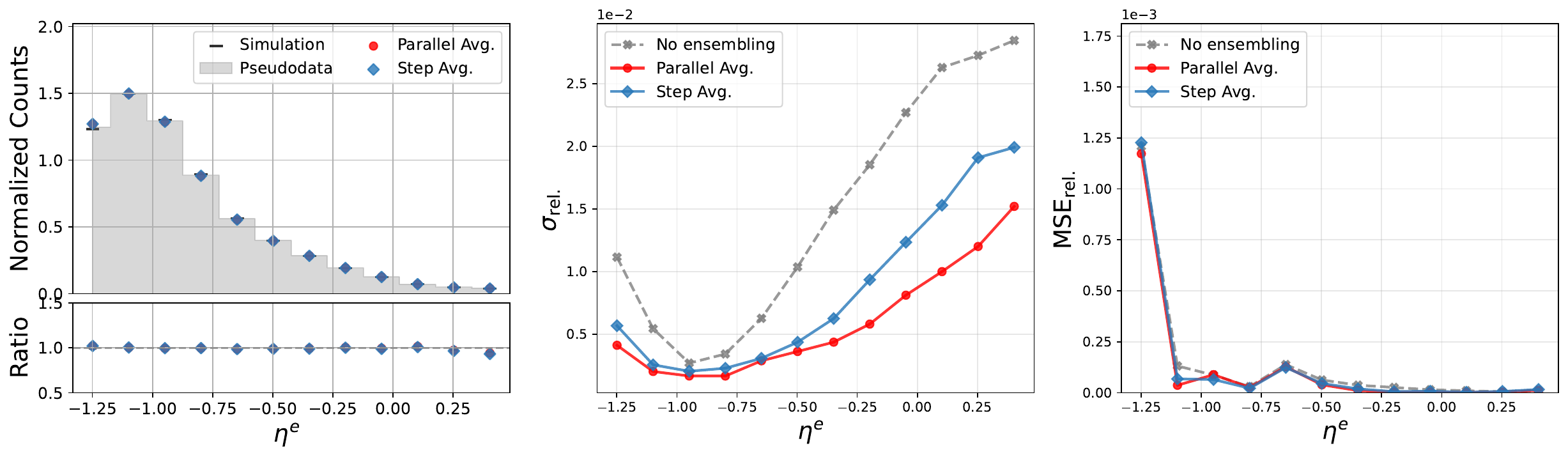}
        \includegraphics[width=0.95\linewidth]{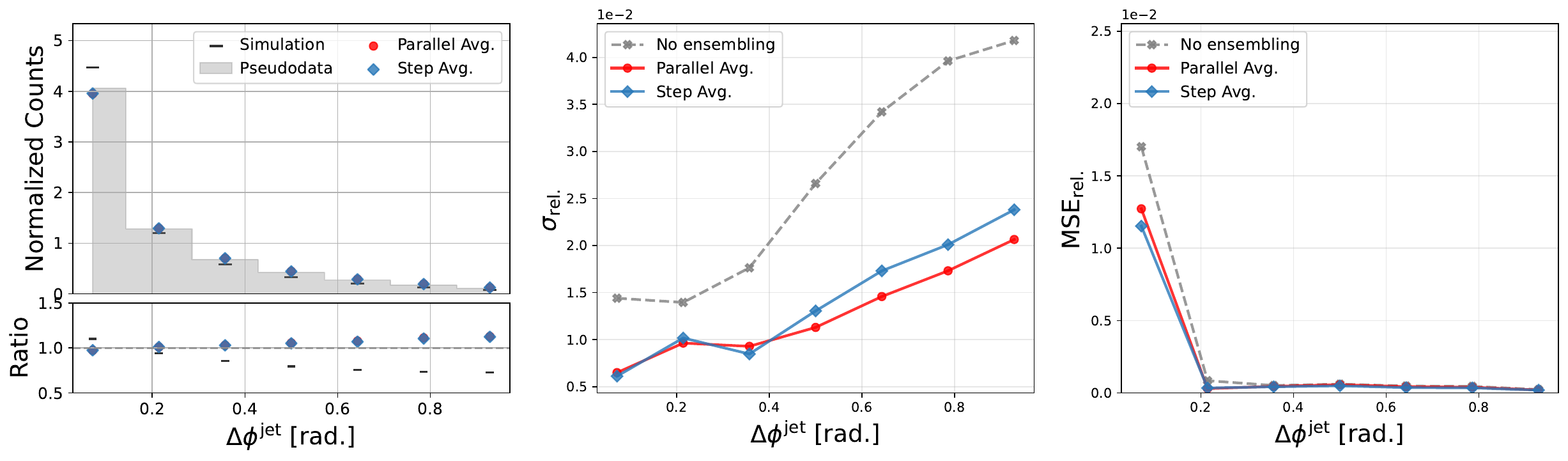}
    \caption{Comparison of $Q^2$ (top), $\eta^e$ (middle), and $\Delta\phi^\text{jet}$ (bottom) for parallel and step ensembling methods with models trained from scratch.  The middle column shows the relative standard deviation of the 10 passes, and the right column show the relative MSE.}
    \label{fig:event_obs}
\end{figure}

Next, we compare parallel and step unfolding without pretraining, using particle-level (Fig.~\ref{fig:part_obs}), jet-level (Fig.~\ref{fig:jet_obs}), and event-level (Fig.~\ref{fig:event_obs}) observables.  At particle-level, peak of $\eta_p-\eta_e$ is set by the asymmetric nature of the proton-electron collisions (approximately 30 GeV electrons and 1 TeV protons) and the $p_T$ spectrum is determined by the $Q^2$ and the fragmentation functions.  In the laboratory frame, most events have at least one jet (to balance the scattered electron) with a steeply falling $p_T$ spectrum.  In deep-inelastic scattering and at HERA energies, most jets are quark jets and tend to have a high value of $\tau_{21}$ relative to jets with two hard-prongs (e.g. from a Lorentz-boosted $W$ boson decaying hadronically).  At the event level, the $Q^2$ values are steeply falling from the $150$ GeV$^2$ threshold.  Most jets are back-to-back with the electron in the transverse plane and so the $\Delta\phi^\text{jet}$ observable instead measures the absolute deviation from $\pi$.  Deviations from $0$ in this observable are due to intrinsic oscillations of the partons within the proton and from initial/final state radiation.

The full phase space is too high-dimensional to probe holistically, but our selection of observables are representative of different aspects of the events.  Across the board, the parallel and step ensembling performance is similar, achieving excellent accuracy and precision.  Overall, the parallel ensembling achieved a better precision with little cost in bias.  This is most prominently displayed by the particle-level obervables, including a nearly factor of two improvement at high $\eta_p-\eta_e$ in the tails of the top row in Fig.~\ref{fig:part_obs} and a similar factor for low $p_T$ particles in the second row of the same figure. The benefits are less clear for jets, but then grow to 10-20\% for the tails of the event-level observables.



Both, the step and parallel, ensembling methods greatly improve the variance of the results when compared to no ensembling, with a smaller improvement to the overall bias of the results. While the step ensembling method resulted in slightly higher variance, the step-ensembled models trained faster, and were therefore used as the baseline for the next study: investigating the effects of pre-training on unfolding performance, as discussed in Sec \ref{sec:pretrain}.

\begin{figure}[H]
    \centering
    \includegraphics[width=0.95\linewidth,trim={4cm 0 4cm 0},clip]{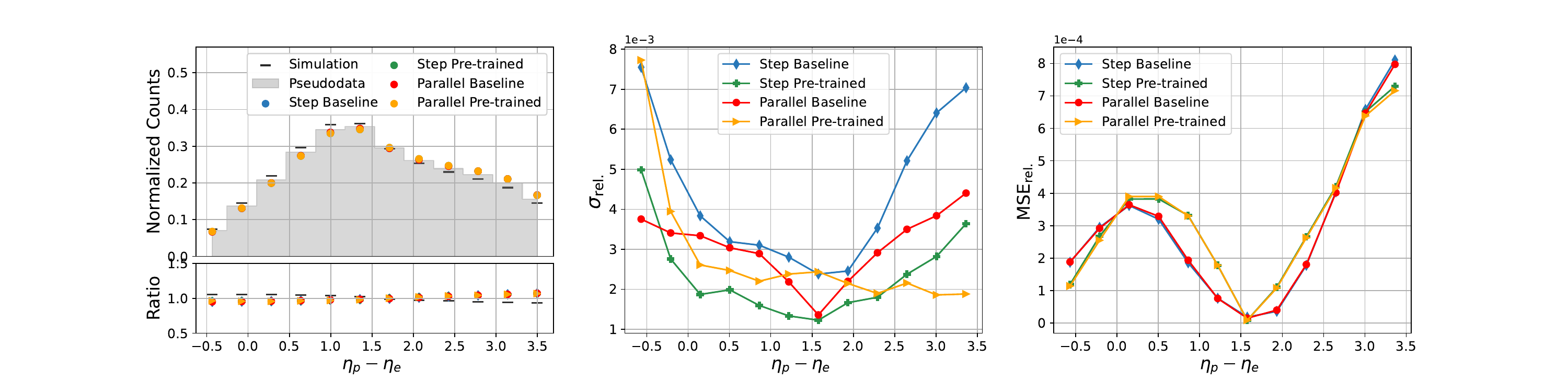}
    \includegraphics[width=0.95\linewidth,trim={4cm 0 4cm 0},clip]{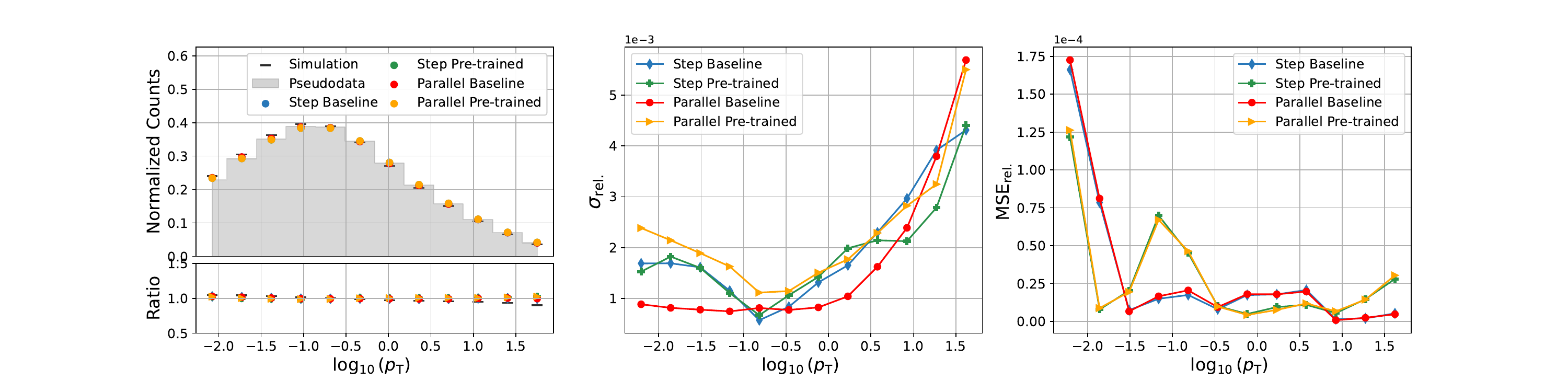}
    \includegraphics[width=0.95\linewidth,trim={4cm 0 4cm 0},clip]{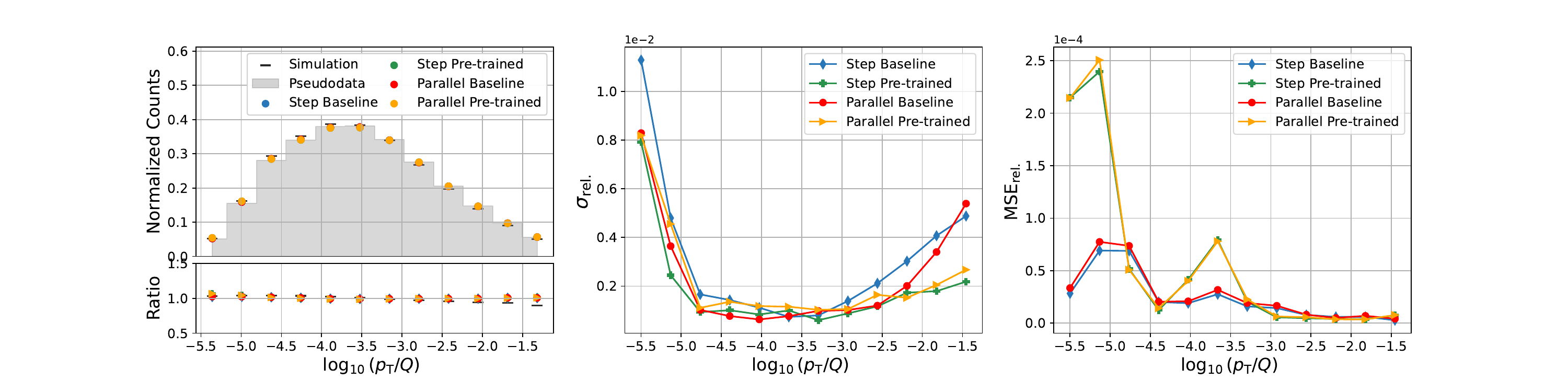}

    \caption{Comparison of relative particle pseudorapidity, $\eta_p-\eta_e$ (top), $\log_{10}(p_\mathrm{T}^\mathrm{part.})$ (middle), and $\log_{10}p_\mathrm{T}^\mathrm{part.}$ (bottom) for pre-trained and from-scratch models. The middle column shows the relative standard deviation of the 10 passes, and the right column show the relative MSE.
    %
    }
    \label{fig:part_PreVsBase}
\end{figure}

\begin{figure}[H]
    \centering
    \includegraphics[width=0.95\linewidth,trim={4cm 0 4cm 0},clip]{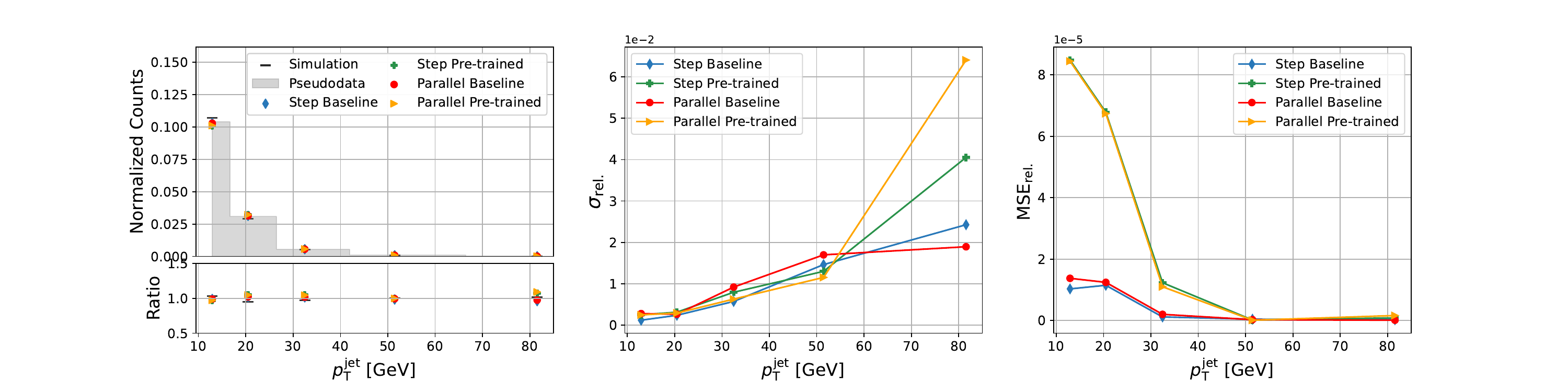}
    \includegraphics[width=0.95\linewidth,trim={4cm 0 4cm 0},clip]{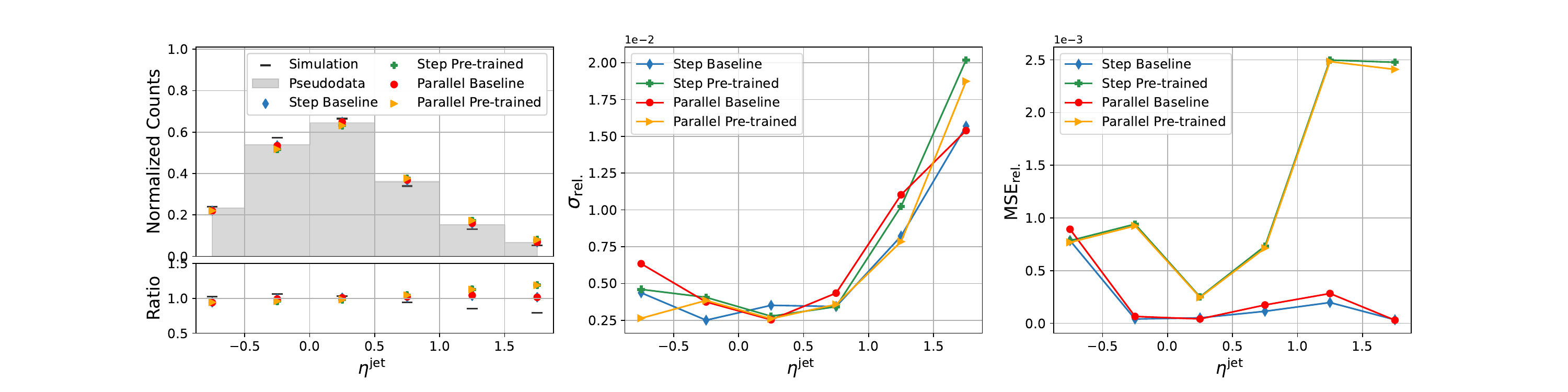}
    \includegraphics[width=0.95\linewidth,trim={4cm 0 4cm 0},clip]{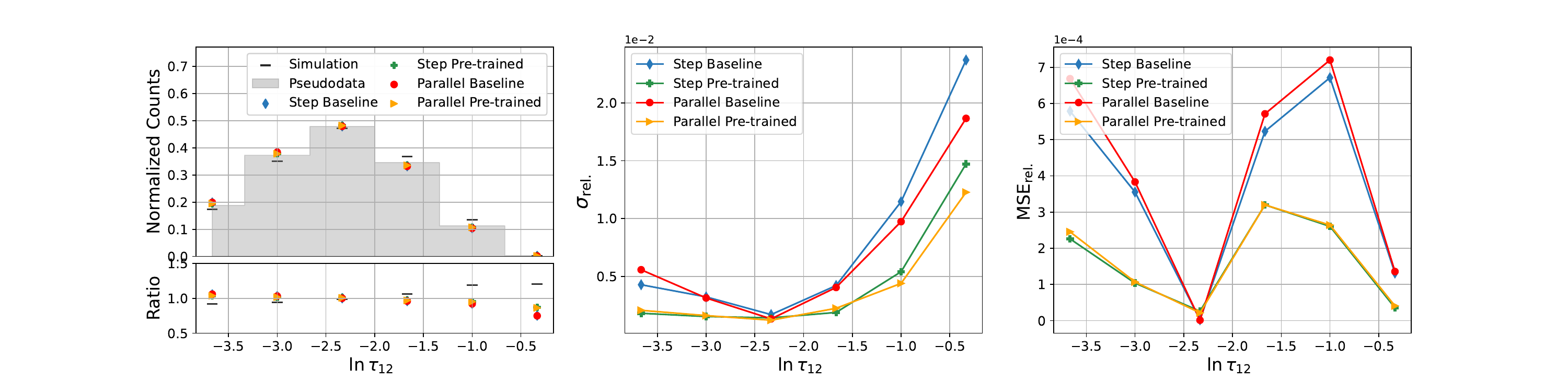}
    \caption{Comparison of $p_\mathrm{T}^\mathrm{jet}$ (top) $\eta^\mathrm{jet}$ (center) and jet $\tau_{21}$ (bottom) for pre-trained and from-scratch models.  The middle column shows the relative standard deviation of the 10 passes, and the right column show the relative MSE.}
    \label{fig:jet_PreVsBase}
\end{figure}

\begin{figure}[H]
    \centering
    \includegraphics[width=0.95\linewidth,trim={4cm 0 4cm 0},clip]{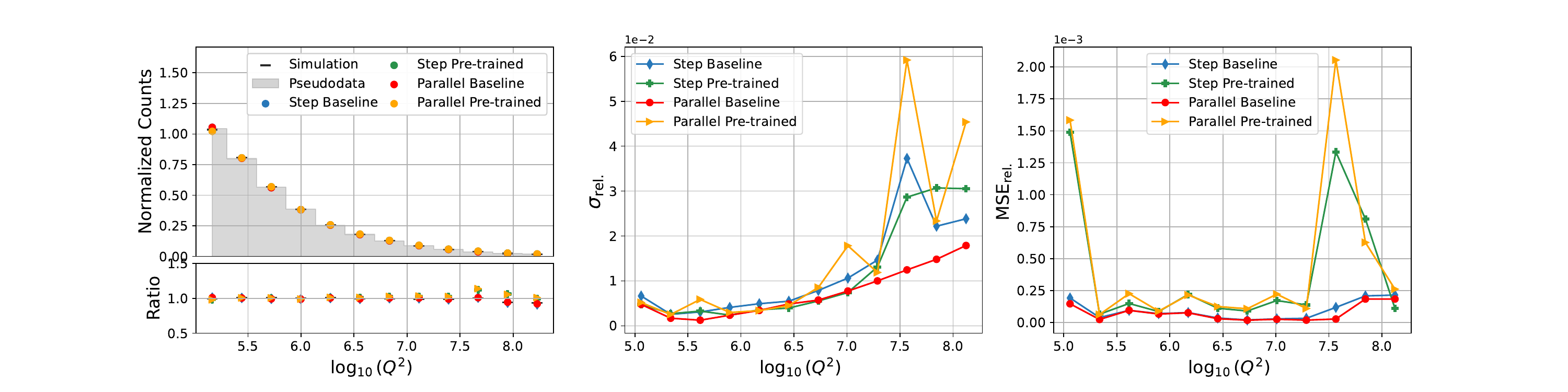}
    \includegraphics[width=0.95\linewidth,trim={4cm 0 4cm 0},clip]{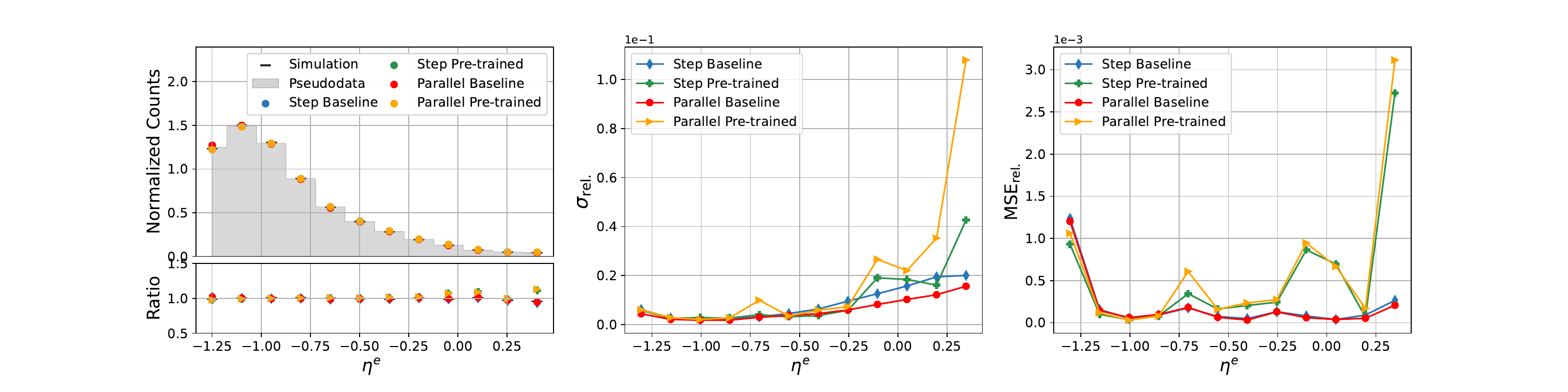}
    \includegraphics[width=0.95\linewidth,trim={4cm 0 4cm 0},clip]{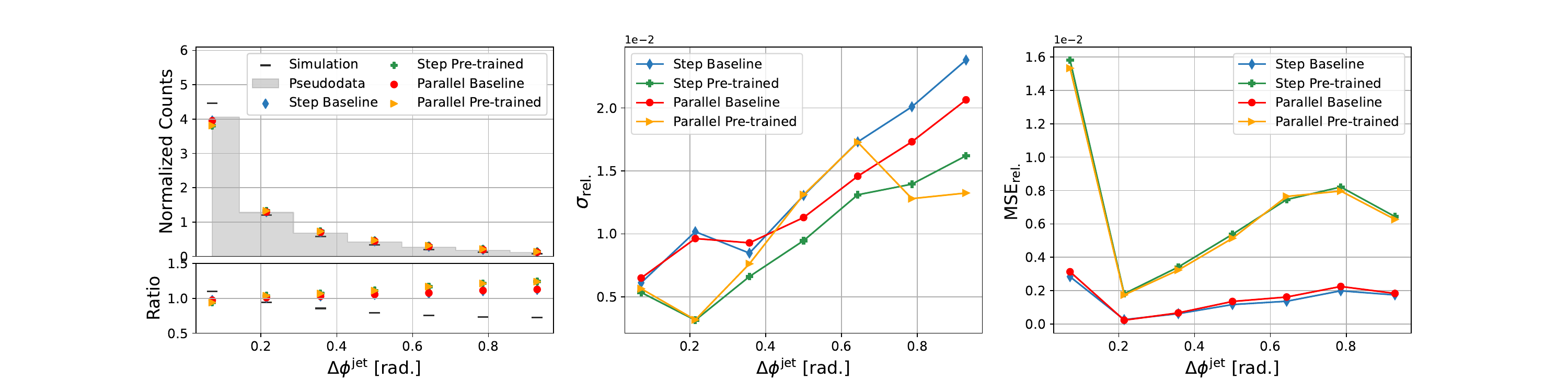}
    \caption{Comparison of $Q^2$ (top), $\eta^e$ (middle), and $\Delta\phi^\text{jet}$ (bottom) for pre-trained and from-scratch models.  The middle column shows the relative standard deviation of the 10 passes, and the right column show the relative MSE.}
    \label{fig:event_PreVsBase}
\end{figure}

Figures \ref{fig:part_PreVsBase} (particle level), \ref{fig:jet_PreVsBase} (jet level), and~\ref{fig:event_PreVsBase} (event level), show a comparison of the baseline step- or parallel-ensembling approach, trained from scratch with randomly initialized weights, and the pre-trainind step or parallel ensembling.  As expected, pre-training generally improves the precision.  For particle pseudorapidity, this improvement is up to 50\% and it is nearly a factor of two in the tails of the $p_T/Q$ distribution.  The largest improvements for jets is for $\tau_{21}$, with up to 50\% reduction in the relative standard deviation in the tails.  While the improved precision is large for $\Delta\phi^\text{jet}$, the benefits are more modest for $Q^2$ and $\eta^e$.  The benefits are less clear for parallel ensembling compared to step ensembling.

The price for this reduction in variance is an increase in bias.  This may be expected because the pre-trained model always starts in the same place and is thus less-able to explore solutions far from this initial guess.  The bias is worst for the jet-level observables (aside from $\tau_{21}$) and also large in the tails of the particle-level and event-level distributions.


In addition to accuracy and precision, another important consideration is training time.  This is especially true for high-dimensional likelihood-ratio estimation problems, where it can take substantial time even on graphical processing units (GPUs) to train a single model.  We generally expect step ensembling to be a little faster than parallel ensembling because each model is a bit better from previous steps.  We also expect pre-training to improve the training time, since the models start closer to a reasonable solution.  These trends are confirmed numerically with training times for a full pass shown in Table \ref{tab:training_times}. The training is done on the Perlmutter Supercomputer~\cite{Perlmutter} using 64 A100 GPUs simultaneously with Horovod~\cite{sergeev2018horovod} package for data distributed training. A local batch size of size 256 is used with model training up to 100 epochs.  The parallel and step models with pre-training use the same pre-trained model; the time to train this model is not shown, but becomes irrelevant when it is reused many times.

\begin{table}[H]
\centering
\begin{tabular}{|l |c|}
\hline
\textbf{Method} & \textbf{Train Time (hrs)} \\
\hline
Parallel Ensemble & 11.62 $\pm$ 0.09 \\
Pre-trained Parallel Ensemble & 8.52 $\pm$ 0.08\\
Step Ensemble     & 8.85 $\pm$ 0.09 \\
Pre-trained Step Ensemble  & 8.26 $\pm$ 0.08 \\
\hline
\end{tabular}
\caption{Training times (in hours) for a full pass of OmniFold trained on 5M simulation and pseudodata samples. The error bars represent the statistical uncertainty from training 10 full passes of OmniFold.}
\label{tab:training_times}
\end{table}



\section{Conclusion and Outlook}
\label{sec:conc}
\label{conclusion}

Mitigating the stochastic nature of neural network training is an important aspect of precision science with neural likelihood ratio estimation.  In this paper, we explore a number of strategies to stabilize the likelihood ratio.  In particular, we study ensembling and pre-training.  As a representative task with many likelihood-ratio estimation sub-tasks, we use unbinned unfolding from simulated deep inelastic scattering.

Both ensembling and pre-training reduce the variance.  Running the full multi-step unfolding procedure and then combining (parallel ensembling) generally led to a lower variance than combining the results one network training at a time (step ensembling) without an increase in the bias.  The parallel ensembling is also easier to implement and trivially parallizable.  It is also possible to change the combination procedure after the fact (mean, median, truncated mean).  The step ensembling is faster, but the difference in speed can be largely eliminated with pre-training.  Pre-training generally reduces the variance for step ensembling, but we found it gave more mixed results for parallel ensembling.  In general, pre-training led to a higher bias.  

Based on these results, our general recommendation is to apply pre-training when limited by the amount of real data, or if the lower variance is worth the cost of higher bias. Additionally, the parallel ensembling method is chosen over the step-ensembling method due to the improved variance and greater flexibility.  

There may still be possibilities for further improving the performance.  In general, there may be a non-trivial interplay between the network architecture, training protocol, and the ensembling.  Going forward, the optimization of a strategy for a particular dataset should include different stabilization strategies as hyperparameters to be explored.  It may also be possible to reduce the bias from pre-training by using multiple pre-trained models instead of just one.  The pre-training speed may also be accelerated by lowering the patience, especially since most of the learning and time (improvement) comes in the first step.  These and other strategies will further enable a precision measurement program for parameters and differential cross sections at colliders and beyond.

\section*{Code and Data availability}
The code in this work can be found in: \url{https://github.com/ViniciusMikuni/H1Unfold/tree/step_ensemble}.

\section*{Acknowledgments}
This material is based upon work supported by the National Science Foundation under Grant No. 2311666.  This research used resources of the National Energy Research Scientific Computing Center, a DOE Office of Science User Facility supported by the Office of Science of the U.S. Department of Energy under Contract No. DE-AC02- 05CH11231 using NERSC awards HEP-ERCAP0021099 and HEP-ERCAP0028249.  We express our thanks to all those involved in securing not only the H1 data but also the software and working environment for long term use, allowing the unique H1 data set to continue to be explored. The transfer from experiment specific to central resources with long term support, including both storage and batch systems, has also been crucial to this enterprise. We therefore also acknowledge the role played by DESY-IT and all people involved during this transition and their future role in the years to come.

\bibliographystyle{unsrt} 
\bibliography{HEPML, other, EEC_ref} 

\end{document}